\documentclass[11pt]{article}
\usepackage{amsfonts}
\usepackage{pstricks}
\usepackage{pst-node}
\usepackage{epsfig}
\usepackage{epsfig}
\usepackage{graphicx,color}

\begin{document}

\def\ba{\begin{eqnarray}}
\def\ea{\end{eqnarray}}
\def\w{\wedge}
\def\d{\mbox{d}}
\def\D{\mbox{D}}

\begin{titlepage}
\title{\bf Symmetric Teleparallel Gravity: {\Large Some exact solutions and spinor couplings}}
\author{ Muzaffer Adak${}^1$\footnote{\tt madak@pau.edu.tr} , \"{O}zcan Sert${}^2$\footnote{\tt osert@pau.edu.tr}
, Mestan Kalay${}^1$\footnote{\tt mkalay@pau.edu.tr} , Murat Sar{\i}${}^2$\footnote{\tt msari@pau.edu.tr} \\ \\
 {\small ${}^1$ Department of Physics, Faculty of Arts and Sciences, Pamukkale University}\\
 {\small 20017 Denizli, Turkey } \\
 {\small ${}^2$ Department of Mathematics, Faculty of Arts and Sciences, Pamukkale University}\\
 {\small 20017 Denizli, Turkey }
  }
  \vskip 1cm
\date{ {06 November 2013}, {\it file CosmicSpeedUpInSTPG21.tex}  }
\maketitle

 \thispagestyle{empty}

\begin{abstract}
 \noindent
In this paper we elaborate on the symmetric teleparallel gravity
(STPG) written in a non-Riemannian spacetime with nonzero
nonmetricity, but zero torsion and zero curvature. Firstly we give
a prescription for obtaining the nonmetricity from the metric in a
peculiar gauge. Then we state that under a novel prescription of
parallel transportation of a tangent vector in this non-Riemannian
geometry the autoparallel curves coincides with those of the
Riemannian spacetimes. Subsequently we represent the symmetric
teleparallel theory of gravity by the most general quadratic and
parity conserving lagrangian with lagrange multipliers for
vanishing torsion and curvature. We show that our lagrangian is
equivalent to the Einstein-Hilbert lagrangian for certain values
of coupling coefficients. Thus we arrive at calculating the field
equations via independent variations. Then we obtain in turn
conformal, spherically symmetric static, cosmological and pp-wave
solutions exactly. Finally we discuss a minimal coupling of a
spin-$1/2$ field to STPG.

\vskip 1cm


\end{abstract}
\end{titlepage}

\section{Introduction}

Einstein's theory of gravity, the so-called general relativity
(GR), is formulated in a Riemannian spacetime. But there are
convincing reasons to go beyond the Riemannian spacetime coming
from both mathematical and physical points of view. Mathematically
both group theoretical \cite{kirsch2005},\cite{boulanger2006} and
gauge theoretical \cite{benn1982},\cite{fhehl1995} approaches
predict non-Riemannian geometries with especially nonmetricity.
Physically unsuccessful affords to quantize gravity with standard
field theoretical methods are longstanding reasons for
modification of GR. Furthermore the recent observational data
received from type-Ia supernovae have showed that the expansion of
the Universe is currently undergoing a period of acceleration
\cite{ariess1998},\cite{perlmutter1999}. Cosmic speed-up can be
explained within general relativity by invoking a mysterious
cosmic fluid with large negative pressure, the so-called dark
energy. Although the simplest possibility for dark energy is a
cosmological constant, unfortunately the estimated value for it is
too large compared with today's value. Consequently GR must be
modified or replaced with new one.

Einstein's equations
 \ba
      (Ric)_a - \frac{1}{2} \mathcal{R} e_a + \Lambda e_a = \kappa \, ^*\!t_a
      \label{eq:EinEqn}
 \ea
are obtained as local extremum of the action $I=\int_M L_0$ where
 \ba
     L_0 = -\frac{1}{2\kappa} R^a{}_b \w ^*\!e_a{}^b + \frac{\Lambda}{\kappa} \, {}^*\!1 + \lambda_a \w T^a +
     \mu_{ab} \w Q^{ab} - K \label{eq:EinHilbertLag}
 \ea
is the Einstein-Hilbert lagrangian 4-form with the cosmological
constant $\Lambda$. Here $\kappa$ is a coupling constant, $t_a =
\delta K / \delta e^a$ are the matter energy-momentum 3-forms,
$\lambda_a$ are the Lagrange multiplier 2-forms constraining
torsion to zero, $\mu_{ab}$ are the Lagrange multiplier 3-forms
constraining nonmetricity to zero. Other notations will be
introduced in the next section. In one approach of modified
gravity models one passes to the non-Riemannian geometries;
$R^a{}_b \neq 0$, $T^a\neq 0$ and $Q_{ab} \neq 0$. A gravity model
developed in such a spacetime is often called the metric-affine
gauge theory of gravity (MAG) \cite{fhehl1995}. Despite a big
degrees of freedom of those models it is not an easy matter to
handle the most general case. Therefore people put some
restrictions on their gravity models, see for example
\cite{fhehl1995},\cite{rtucker1995},\cite{madak2010},\cite{tdereli1996},\cite{dereli2002},\cite{pbaekler2011}
and the references therein. In this work we adhere the Yang-Mills
manner for a gravitational lagrangian, that is, we consider the
most general nonmetricity quadratic and parity conserving
lagrangian 4-form
 \ba
     L = \sum_{I=1}^4 k_I Q_{ab} \w ^*\!{}^{(I)}\!Q^{ab}
     + k_5 \left({}^{(3)}\!Q_{ab} \w e^b\right) \w ^*\!\left({}^{(4)}\!Q^{ac} \w e_c\right)
     + \Lambda \, ^*\!1 + \lambda_a \w T^a +
     R^a{}_b \w \rho^b{}_a - K \label{eq:STPGlagrangian}
 \ea
where $k_I$s are coupling constants, ${}^{(I)}Q^{ab}$ are
irreducible decompositions of nonmetricity
\cite{madak2010},\cite{madak2006}, $\Lambda$ is the cosmological
constant, $K$ is the matter Lagrangian 4-form, $\lambda_a$ are the
Lagrange multiplier 2-forms for zero-torsion, $\rho^b{}_a$ are the
Lagrange multiplier 2-forms for teleparallelism. In other words,
we have simply replaced the left hand side of (\ref{eq:EinEqn})
with a new one written in terms of only nonmetricity. This model
is called symmetric teleparallel gravity (STPG)
\cite{jmnester1999}. Thereby metric $g_{\alpha \beta}$ is seen as
gauge potential and the corresponding gauge field strength is
nonmetricity $Q_{\alpha \beta} \sim \D g_{\alpha \beta}$ that
satisfies the Bianchi identity $\D Q_{\alpha \beta} =0$.

There is an another teleparallel approach to gravity, the
so-called teleparallel gravity (TPG), formulated only with
torsion. Although TPG has attracted a lot of attention in the
literature, STPG is infancy. One can consult, for example, the
references  \cite{obukhov2003},\cite{maluf2012} to start reading
the wide literature on TPG. On the other hand, a few physically
interesting analysis and solutions of STPG can be found in
\cite{madak2006},\cite{jmnester1999},\cite{madak2005},\cite{madak2008}.
Our wish here is to fill in a bit this big gap at STPG. Our other
motivation is to gain new insights on the nometricity. Recently it
has been related with the breaking of the Weyl symmetry
\cite{rtucker1995}, with the spin-3 gravity field
\cite{baekler2006} and used in the solar neutrino mixing problem
\cite{adak2004}.

The outline of the paper is as follows. In Section
\ref{sec:MathPreli} after introducing the non-Riemannian geometry
and then symmetric teleparallel geometry we define a gauge fixing
procedure and a deliberative parallel transportation rule for a
tangent vector. From now on we use the computer algebra system
REDUCE and its package EXCALC intensively
\cite{hearn1993},\cite{schrufer1994}. In Section
\ref{sec:fieldeqns} we present a systematic treatment of STPG
through a lagrange 4-form quadratic in the nonmetricity. Here we
show explicitly that for certain values of coupling coefficients
our lagrangian corresponds to the Einstein-Hilbert lagrangian in
Riemannian spacetime. Then the field equations are obtained
consistently by independent variations. By solving $\D \lambda_a$
from the connection varied equation and then by inserting it into
the coframe varied equation we write down the dynamical field
equation (\ref{AlanDenk6}) in a naive way. In the next four
sections we obtain some classes of exact solutions in the type of
conformal, spherically symmetric static, cosmological and pp-wave,
respectively. Finally we analyze a minimal coupling of a Dirac
spinor to STPG. Section \ref{sec:result} is devoted for the
concluding remarks.

\section{ Mathematical preliminaries} \label{sec:MathPreli}

Spacetime is denoted by the triple $\{M,g,\nabla \}$ where $M$ is
differentiable and orientable 4-dimensional manifold, $g=g_{\alpha
\beta}e^\alpha \otimes e^\beta$ is the metric tensor, and $\nabla$
represents connection. $\{ X_\alpha \}$ denotes the basis vector
set of tangent space $T_p(M)$ at the point $p$ of manifold.
Similarly, $\{ e^\alpha \}$ denotes the basis co-vector set of
cotangent space $T_p^*(M)$ at the point $p$ of manifold. We show
duality as
 \begin{eqnarray}
          e^\alpha(X_\beta)\equiv
          \iota_{X_\beta}(e^\alpha)=\delta^\alpha_\beta
 \end{eqnarray}
where $ \delta^\alpha_\beta$ is Kronecker symbol and
$\iota_{X_\beta} \equiv \iota_\beta$ is the interior product. Full
connection is determined by local 1-forms, $\omega^\alpha{}_\beta$
as follows:
 \ba
    \nabla_{X_\alpha}X_\beta = \omega^\gamma{}_\beta(X_\alpha) X_\gamma \, .
 \ea
From now on we call $e^\alpha$ basis 1-forms and $
\omega^\alpha{}_\beta$ full connection 1-forms. Thus, under a
general coordinate transformation while basis 1-forms transform as
 \ba
   e^{\alpha'}= h^{\alpha'}{}_\alpha e^\alpha
 \ea
full connection 1-forms transform in the following manner
 \ba
    \omega^{\alpha'}{}_{\beta'}= h^{\alpha'}{}_\alpha
    \omega^{\alpha}{}_\beta h^{\beta}{}_{\beta'} +
     h^{\alpha'}{}_\alpha \d h^{\alpha}{}_{\beta'}
     \label{ConTransform}
 \ea
where $h^{\alpha'}{}_\alpha$ are transformation elements. The following
shorthand notation for exterior multiplications is adhered
$e^\alpha \wedge e^\beta \wedge \cdots \equiv e^{\alpha \beta
\cdots }$. A geometry is determined by the Cartan structure equations
 \ba
     Q_{\alpha \beta} &:=& - \frac{1}{2} \D g_{\alpha \beta}
                 = \frac{1}{2} (-\d g_{\alpha \beta} + \omega_{\alpha \beta}+\omega_{\beta \alpha}) \, , \label{nonmet}\\
     T^\alpha &:=& \D e^\alpha = \d e^\alpha + {\omega^\alpha}_\beta \wedge e^\beta \, , \label{tors}\\
     {R^\alpha}_\beta &:=& \D {\omega^\alpha}_\beta := \d {\omega^\alpha}_\beta
                  + {\omega^\alpha}_\gamma \wedge {\omega^\gamma}_\beta \, , \label{curva}
 \ea
where $Q_{\alpha \beta}$ are the nonmetricity 1-forms,
$T^\alpha$ are the torsion 2-forms and $ {R^\alpha}_\beta$
are the curvature 2-forms. $ \d$ and $ \D$ denote the
exterior derivative and the covariant exterior derivative,
respectively. These tensors satisfy the Bianchi identities
 \ba
       \D Q_{\alpha \beta} &=& \frac{1}{2} ( R_{\alpha \beta} +R_{\beta \alpha}) \; , \label{bianc:0} \\
       \D T^\alpha    &=& {R^\alpha}_\beta \wedge e^\beta \; , \label{bianc:1} \\
       \D {R^\alpha}_\beta &=& 0  \; . \label{bianc:2}
 \ea
In general, full connection 1-forms can be decomposed as follows
\cite{fhehl1995},\cite{rtucker1995},\cite{madak2010}:
 \ba
     {\omega^\alpha}_\beta =  \underbrace{(g^{\alpha \gamma} \d g_{\gamma_\beta} + {p^\alpha}_\beta)/2 +
     {\widetilde{\omega}}^\alpha{}_\beta}_{Riemannian}
     + \underbrace{{K^\alpha}_\beta}_{Torsional} + \underbrace{{q^\alpha}_\beta
     + {Q^\alpha}_\beta}_{Nonmetrical}   \label{connect:dec}
 \ea
where Levi-Civita connection 1-forms
$\widetilde{\omega}^\alpha{}_\beta$
 \ba
     \widetilde{\omega}^\alpha{}_\beta \wedge e^\beta = -\d e^\alpha  \; , \label{LevCiv}
 \ea
contortion 1-forms $K^\alpha{}_\beta$
 \ba
   {K^\alpha}_\beta  \wedge e^\beta = T^\alpha  \; , \label{contort}
 \ea
and anti-symmetric 1-forms
 \ba
    & & p_{\alpha \beta} = -( \iota_\alpha  \d g_{\beta \gamma } ) e^\gamma + ( \iota_\beta \d g_{\alpha
   \gamma})  e^\gamma \, , \label{p:ab} \\
   & & q_{\alpha \beta} = -( \iota_\alpha  Q_{\beta \gamma } ) e^\gamma + ( \iota_\beta Q_{\alpha \gamma})
    e^\gamma  \; . \label{q:ab}
 \ea
This decomposition is self-consistent. To see that one just has to
multiply (\ref{connect:dec}) by $\w e^\beta$ and to use above
definitions, where special care is needed to raise and
lower an index in front of both $\d$ and $\D$. Symmetric part
of (\ref{connect:dec}) comes from (\ref{nonmet})
 \ba
  \omega_{(\alpha \beta)} = Q_{\alpha \beta } + \frac{1}{2}\d g_{\alpha \beta } \label{connect:sym}
 \ea
and the remaining is anti-symmetric
 \ba
  \omega_{[\alpha \beta]} = \frac{1}{2} p_{\alpha \beta} + \widetilde{\omega}_{\alpha \beta}
  + K_{\alpha \beta} + q_{\alpha \beta} \; .   \label{connect:ansym}
 \ea
We denote symmetrization and anti-symmetrization by
$\omega_{(\alpha \beta)}:=\frac{1}{2}\left(\omega_{\alpha \beta} +
\omega_{\beta \alpha}\right)$ and $\omega_{[\alpha
\beta]}:=\frac{1}{2}\left(\omega_{\alpha \beta} - \omega_{\beta
\alpha}\right)$, respectively.

In general, the triple of nonmetricity, torsion and curvature
classifies a geometry. If $Q_{\alpha \beta}=0$, $T^\alpha \neq 0$
and $R^\alpha{}_\beta \neq 0$, we have a  metric compatible
connection in the Riemann-Cartan (RC) geometry. If $Q_{\alpha
\beta}=0$, $T^\alpha =0$ and $R^\alpha{}_\beta \neq 0$, then we
have the Levi-Civita connection of a Riemann geometry. If
$Q_{\alpha \beta} = 0$, $T^\alpha \neq 0$ and $R^\alpha{}_\beta
=0$, it is called the Weitzenb\"{o}ck geometry. Here we work in
the symmetric teleparallel geometry defined by $Q_{\alpha \beta}
\neq 0$, $T^\alpha =0$ and $R^\alpha{}_\beta =0$.

It is always possible to choose orthonormal basis 1-forms which we
denote $\{ e^a(x) \}$. Then the metric turns out to be
$g=\eta_{ab}e^a(x) \otimes e^b(x)$ where $\eta_{ab} =
\mbox{diag}(-1,+1,+1,+1)$. In this case, the orientation of the
manifold can be fixed by the Hodge star ${}^*\!1=e^{0123}$ and
also the connection 1-forms decompose according to
 \ba
   \omega^a{}_b =  \widetilde{\omega}^a{}_b + K^a{}_b + q^a{}_b
   + Q^a{}_b \; . \label{connect:decortonormal}
 \ea
After this point while the Greek indices figure holonomic
(coordinate) indices, the Latin indices are anholonomic
(orthonormal) ones.

\subsection*{A gauge fixing}

We will be working in the orthonormal frame through the paper. But
we can not solve equations $T^a=\d e^a + \omega^a{}_b \w e^b=0$
and $R^a{}_b = \d \omega^a{}_b + \omega^a{}_c \w \omega^c{}_b=0$
together for $\omega^a{}_b$. This is opposed to the Riemannian
case in which one can solve $\omega^a{}_b$ uniquely from
$Q_{ab}=(\omega_{ab}+\omega_{ba})/2=0$ and $T^a=\d e^a +
\omega^a{}_b \w e^b=0$. Thus one method may be to guess suitable
coframe and connection such that zero-torsion and zero-curvature
are satisfied \cite{madak2006}. In this work, instead of that
approach, we prefer to develop a general recipe for treating the
problem. In our recipe we start calculation in the coordinate
frame in which the metric is $ g= g_{\alpha \beta}(x) \d x^\alpha
\otimes \d x^\beta $ and the coframe is  $ e^\alpha = \d x^\alpha
$ where $\{ x^\alpha \}$ is a local coordinate system.  Now we
choose a special coordinate system where $\omega^\alpha{}_\beta
=0$. This may be seen as a gauge fixing as well. In this gauge (or
coordinate), two conditions, $T^\alpha=0$ and
$R^\alpha{}_\beta=0$, are met automatically and we obtain non-zero
nonmetricity $Q_{\alpha \beta} = - \frac{1}{2} \d g_{\alpha \beta}
\neq 0$, see the equations (\ref{nonmet})-(\ref{curva}). Then we
need to go back the orthonormal frame via local vierbein
$h^a{}_\alpha$ where $e^a = h^a{}_\alpha \d x^\alpha $.

Before diving transformation details, it is worthwhile to remind a
well-known fact. Since both curvature and torsion are tensors,
they are invariant under coordinate transformations by definition.
That is, it is not possible to make them nonzero by gauge
transformation or fixing if they are zero in some other basis. The
story, however, is completely different for connection since it is
not tensor by definition. In other words, it is possible to make
it nonzero by a gauge transformation even if it is zero in some
other gauge. Accordingly, in the orthonormal frame, one calculates
full connection $\omega^a{}_b = h^a{}_\alpha \omega^\alpha{}_\beta
h^\beta{}_b + h^a{}_\alpha \d h^\alpha{}_b = h^a{}_\alpha \d
h^\alpha{}_b \neq 0 $, full curvature $ R^a{}_b = h^a{}_\alpha
R^\alpha{}_\beta h^\beta{}_b = 0$, torsion $ T^a = h^a{}_\alpha
T^\alpha =0 $ and nonmetricity $Q^a{}_b = h^a{}_\alpha
Q^\alpha{}_\beta h^\beta{}_b \neq 0 $. Finally we have checked
that this result corresponds to $\widetilde{\omega}_{ab} + q_{ab}
=0$ (together with $K_{ab}=0$ since $T^a=0$)  which means
$\omega_{ab} = Q_{ab}$ via (\ref{connect:decortonormal}).

Here since our method crucially relies on a very strong condition
for the full connection $\omega^\alpha{}_\beta$ to vanish in the
coordinate basis which we describe  as a gauge fixing, it is
natural to ask; does such a gauge always exist or what are the
conditions for it? According to (\ref{ConTransform}), the
components of the full connection transform inhomogeneously under
a local linear gauge transformation. So, even when one starts with
a non-zero full connection, $\omega^\alpha{}_\beta$, a vanishing
full connection, $\omega^{\alpha'}{}_{\beta'}$, is reached under
the condition $ h^{\alpha'}{}_\alpha \d h^{\alpha}{}_{\beta'} +
h^{\alpha'}{}_\alpha \omega^{\alpha}{}_\beta
h^{\beta}{}_{\beta'}=0$ where the transformation elements $
h^{\alpha'}{}_\alpha$ define a diffeomorphism. When one considers
a coordinate transformation $x^{\alpha'}=x^{\alpha'}(x^\alpha)$,
it is realized that ${h^{\alpha'}}_\alpha = \partial
x^{\alpha'}/\partial x^\alpha $ and ${h^\alpha}_{\alpha'} =
\partial x^\alpha/\partial x^{\alpha'}$, and then
${h^\alpha}_{\alpha'} {h^{\alpha'}}_\beta = \delta^\alpha_\beta$.
Consequently, the condition above contains a system of the second
order partial differential equations as the following
  \ba
     \frac{\partial^2 x^\alpha}{\partial x^{\alpha'} \partial
     x^{\beta'}} + {\omega^\alpha}_{\beta \alpha'} \frac{\partial x^\beta}{\partial
     x^{\beta'}}=0 \; . \nonumber
  \ea
Thus we can conclude that, in the context of STPG-TPG, such a
gauge always exists, since the integrability conditions of the
above set of equations imply that all components of the curvature
two-form must vanish.

 \subsection*{Autoparallel curves in the symmetric teleparallel geometry}

Here it seems that we have a problem concerning the trajectory of
a test particle. In this geometry trajectories are given by
autoparallel curves. We know that they are basically obtained by
the notion of parallel propagation of tangent vectors to them. Now
we want to review very basic concepts on those issues.

In the Riemann geometry one requires that the tangent vector,
$V^\alpha = dx^\alpha / d\tau$, to the autoparallel curve,
$x^\alpha (\tau)$ with the parameter $\tau$, points in the same
direction as itself when parallel propagated, and demands that it
maintains the same length,
 \ba
    \D V^\alpha =0 \; .
 \ea
When we adhere this definition directly in symmetric teleparallel
geometry together with the gauge $\omega^\alpha{}_\beta=0$ we
obtain an autoparallel equation
 \[
     \d V^\alpha + \omega^\alpha{}_\beta V^\beta=0 \quad \Rightarrow \quad
     \d V^\alpha =0 \quad \Rightarrow \quad \frac{d^2x^\alpha}{d
     \tau^2}=0
 \]
whose solution is simply a straight line. This means that all free
test particles, e.g. planets around the Sun, move on a straight
line. Of course, this is not the case. Therefore, we can not use
this rule of parallel translation. This is the problem mentioned
above and has to be amended.

Thus, since intuitively autoparallel curves are ``those as
straight as possible" we may not demand the vector to keep the
same length \cite{rwald1984} during parallel propagation in STPG.
Besides, it is known that nonmetricity is related to length and
angle standards at parallel transportation
\cite{fhehl1995},\cite{madak2010}. Therefore we prescribe a novel
rule for the parallel propagation of a tangent vector
\cite{madak2011}
  \ba
     \D V^\alpha = (aQ^\alpha{}_\beta + bq^\alpha{}_\beta )
     V^\beta + c Q V^\alpha \label{partrans}
  \ea
where $a,b,c$ are arbitrary constants and $Q=Q^\alpha{}_\alpha$.
Now all trajectories do not need to be straight lines. Since it is
a new problem to determine $a,b,c$, for example, for our solar
system, we leave that for our future project. Nevertheless, we
want to remark one point. When it is chosen $a=b=1$ and $c=0$ in
the gauge $e^\alpha=\d x^\alpha$ and $\omega^\alpha{}_\beta=0$,
the autoparallel curve equation is obtained as
  \ba
  \frac{d^2 x^\alpha}{d\tau^2}
  + \frac{1}{2} g^{\alpha \beta} \left(\partial_\gamma g_{\beta \delta} + \partial_\delta g_{\beta \gamma}
  - \partial_\beta g_{\gamma \delta} \right) \frac{d x^\gamma}{d \tau} \frac{d x^\delta}{d \tau}=0 \, . \label{eq:autoparallel}
 \ea
This is the same as the geodesic equation of the Riemann geometry.

 \section{Field equations of STPG} \label{sec:fieldeqns}

With the assignments
 \ba
    \alpha_a = Q_{ab} \w e^b \, , \quad \beta_a = \iota^b Q_{ab}
    \, , \quad Q=\eta_{ab}Q^{ab} \, , \quad \gamma_a = \iota_a Q
 \ea
the lagrangian (\ref{eq:STPGlagrangian}) is equivalent to the
following
 \ba
     L &=& c_1 Q_{ab} \w ^*\!Q^{ab} + c_2 \alpha_a \w ^*\! \alpha^a + c_3 \beta_a \w ^*\! \beta^a
     + c_4 Q \w ^*\!Q + c_5 \beta_a \w ^*\! \gamma^a \nonumber \\
     & & \quad \quad + \Lambda ^*\!1 + \lambda_a \w T^a +
     R^a{}_b \w \rho^b{}_a - K \label{eq:STPGlagrangian1}
 \ea
under the redefinitions of the coupling coefficients, see \cite{madak2006}
 \ba
 c_1 &=& k_1 \, ,\nonumber \\
 c_2 &=& 2(-k_1 + k_2)/3 \, ,\nonumber \\
 c_3 &=& 2(-k_1 - k_2 + 2k_3)/9 \, , \nonumber \\
 c_4 &=& (-8k_1 - 32k_2 + 4k_3 + 36 k_4 +9k_5)/144  \, , \nonumber \\
 c_5 &=& (-8k_1 + 16k_2 - 8k_3 -9k_5)/36  \, .
 \ea
Since an odd number of Hodge stars occurring in a Lagrangian
warrants its parity conservation, the Lagrangian
(\ref{eq:STPGlagrangian}) or (\ref{eq:STPGlagrangian1}) is the
most general nonmetricity quadratic and parity conserving
lagrangian. Although in the first order formalism we are using,
higher order terms in lagrangian would preserve the second order
of the field equations, the quasilinearity of the field equations
would be damaged and then the Cauchy problem may be ill-posed.
Therefore we consider a lagrangian not higher than nonmetricity
squared.

Certain values of $c_i$s correspond to GR. For determining these
values firstly the full curvature 2-form is decomposed as
Riemannian curvature plus non-Riemannian terms by substituting
(\ref{connect:decortonormal}) (with $T^a=0$) into $R^a{}_b$;
 \ba
    R^a{}_b = \widetilde{R}^a{}_b + \widetilde{\D} \Omega^a{}_b + \Omega^a{}_c \w \Omega^c{}_b
 \ea
where $\Omega_{ab} = Q_{ab} + q_{ab}$. Riemannian quantities are
marked by a tilde. Then the Einstein-Hilbert lagrangian 4-form is
formed
 \ba
    R^a{}_b \w ^*\!e_a{}^b = \widetilde{R}^a{}_b \w ^*\!e_a{}^b
    + \left( \widetilde{\D} \Omega^a{}_b\right) \w ^*\! e_a{}^b
    + \Omega^a{}_c \w \Omega^c{}_b \w ^*\! e_a{}^b \, .
 \ea
Here since $\widetilde{\D} e_a{}^b =0$, the second term is an
exact form, i.e. $\left( \widetilde{\D} \Omega^a{}_b \right) \w
^*\! e_a{}^b = \d \left(\Omega^a{}_b \w ^*\! e_a{}^b \right)$.
Finally, because of vanishing of the full curvature 2-form in this
geometry the left hand is zero. Correspondingly, it may be written
 \ba
    -\frac{1}{2} \widetilde{R}^a{}_b \w ^*\!e_a{}^b = \frac{1}{2} \Omega^a{}_c \w \Omega^c{}_b \w ^*\!e_a{}^b
 \ea
from which one computes the GR-equivalent values of $c_i$s by
comparing with (\ref{eq:STPGlagrangian1})
 \ba
  c_1=-1/2, \;\; c_2=1 , \;\; c_3=0 , \;\; c_4=-1/2 , \;\; c_5=1 \, . \label{GRcs}
 \ea
The advatage of this result is that we can classify the solutions to be obtained as Einsteinian or not.

Now we vary the lagrangian (\ref{eq:STPGlagrangian1}) to obtain
the field equations by noticing $\delta Q_{ab} = \delta
\omega_{(ab)}$ and $(\delta \iota^a)e^b = - \iota^a(\delta e^b)$,
and by using the general formula
 \ba
   \delta(A \w { }^*\!B) = \delta A \wedge { }^*\!B + \delta B \w {
   }^*\!A - \delta e^a \w \left[ (\iota_a B) \w  { }^*\!A - (-1)^{p} A \w (\iota_a {}^*\!B) \right]
 \ea
where $A$ and $B$ are any $p$-forms on $n$-dimensional manifold.
Thus the variations with respect to $\lambda_a$, $\rho^b{}_a$,
$e^a$ and $\omega^b{}_a$ yield the following field equations,
respectively
 \ba
  T^a=0 \, , \quad \quad  R^a{}_b =0 \, ,\\
  \tau_a + \Lambda \,^*\!e_a + \D \lambda_a = t_a \, , \label{CofrEqn} \\
  \Sigma^a{}_b +  e^a \w \lambda_b + \D \rho^a{}_b = \ell^a{}_b \, , \label{ConnEqn}
 \ea
where $t_a = \frac{\delta K}{\delta e^a}$ are the energy-momentum
3-forms of matter, $\ell^a{}_b = \frac{\delta K}{\delta
\omega^b{}_a}$ are the angular momentum 3-forms of matter,
$\tau_a$ are the energy-momentum 3-forms of nonmetricity
 \ba
  \tau_a&=& c_1[\iota_a (Q_{bc} \w { }^*\!Q^{bc}) -2(\iota_a Q_{bc}) \wedge ({ }^*\!Q^{bc})]
         + c_2[\iota_a (\alpha_b \w { }^*\!\alpha^b)-2(\iota_a Q_{bc}) \wedge (e^c \w{ }^*\!\alpha^b)] \nonumber \\
   & & +c_3[\iota_a (\beta_b \w { }^*\!\beta^b)-2(\iota_a Q_{bc}) \wedge (\iota^c {}^*\!\beta^b) ]
    +c_4 [\iota_a (Q \w { }^*\!Q) -2(\iota_a Q_{bc}) \wedge (\eta^{bc}{ }^*\!Q)] \nonumber \\
   & &+c_5[ \iota_a (\beta_b \w { }^*\!\gamma^b) - (\iota_a Q_{bc}) \wedge ( \eta^{bc} \iota_d
   {}^*\!\beta^d + \iota^c {}^*\!\gamma^b )] \label{tauA}
 \ea
and $\Sigma^a{}_b$ are the angular momentum 3-forms of
nonmetricity
 \ba
    \Sigma^{ab} = 2c_1 \, ^*\!Q^{ab} + 2c_2 e^{(a} \w ^*\!\alpha^{b)}
    +2c_3 \iota^{\!(a} \, { }^*\!\beta^{b)} + 2c_4 \eta^{ab} \, {}^*\!Q
    + c_5\left( \eta^{ab} \, \iota_c{}^*\!\beta^c + \iota^{\!(a} \, {}^*\!\gamma^{b)} \right) \, .
 \ea
Let $t_a$ and $\ell^a{}_b$ be known. Then we count the number of
the field equations: 24 from $T^a=0$ plus 96 from $R^a{}_b=0$ plus
16 from (\ref{CofrEqn}) plus 64 from (\ref{ConnEqn}) that add to
200. On the other hand, the number of the unknowns is 200 as well:
16 for $e^a$ plus 64 for $\omega^a{}_b$ plus 24 for $\lambda_a$
plus 96 for $\rho^a{}_b$. Thus the system is closed. In principle,
the Lagrangian multipliers are solved from (\ref{ConnEqn}) and the
result is substituted into (\ref{CofrEqn}). We notice that what we
need in the equation (\ref{CofrEqn}) is only $\D \lambda_a$ rather
than $\lambda_a$ or $\rho^a{}_b$. Thus, by multiplying
(\ref{ConnEqn}) first by $\D$ and then by $\iota_a$ we obtain
 \ba
   \D \lambda_a = \iota _b \D (\Sigma^b{}_a - \ell^b{}_a)
 \ea
where we used $\D e^a = T^a=0$ and $\D^2 \rho^a{}_b = R^a{}_c \w
\rho^c{}_b - R^c{}_b \w \rho^a{}_c=0$. Substitution of this into
(\ref{CofrEqn}) yields the dynamical field equation of STPG
  \ba
    \iota_b \D \Sigma^b{}_a + \tau_a + \Lambda \, {}^*\!e_a  = t_a + \iota_b \D \ell^b{}_a  \; .  \label{AlanDenk6}
 \ea

 \section{Conformal solution} \label{sec.conform}

We consider the metric
 \ba
     g = \exp(2\Phi) \eta_{\alpha \beta} \d x^\alpha \otimes \d x^\beta
 \ea
where $\eta_{\alpha \beta}=\mbox{diag}(-1,+1,+1,+1)$ and
$\Phi=\Phi(t,x,y,z)$ is the conformal function in the cartesian
coordinates, $x^\alpha =(t,x,y,z)$. In the coordinate frame, i.e.
$e^\alpha = \d x^\alpha$, we choose such a gauge that
$\omega^\alpha{}_\beta =0$ and then calculate the nonmetricity via
$Q_{\alpha \beta} = -\frac{1}{2} \d g_{\alpha \beta}$ as
 \ba
   Q_{\alpha \beta} = -\eta_{\alpha \beta} \exp(2\Phi) \d \Phi \, .
 \ea
Now we pass to the orthonormal frame
 \ba
    g=\eta_{ab} e^a \otimes e^b \quad \mbox{with} \quad e^a = \exp(\Phi) \d x^a \, .
 \ea
Accordingly the nonmetricity in the orthonormal frame can be
computed
 \ba
   Q_{ab} = -\eta_{ab} \d \Phi \, .
 \ea
Here we checked that $\widetilde{\omega}_{ab} + q_{ab}=0$. After
substituting this into the equation (\ref{AlanDenk6}) with $t_a=0$
and $\ell^a{}_b=0$ we see that the conformal function is arbitrary
under the constraints
 \ba
    4c_1 + 5c_2 - c_3 + 16c_4=0 \, , \quad 2(c_1 + c_2 + 4c_4) + c_5
    =0 \label{conformconst}
 \ea
with $\Lambda =0$.  We also remark that GR-values of $c_i$s
(\ref{GRcs}) do not satisfy these constraints.

 \section{Spherically symmetric static solutions}

We write the following metric in spherical polar coordinates
$(r,\theta,\phi)$
 \ba
    g = -F(r)^2 \d t^2 + G(r)^2 \d r^2 + r^2(\d \theta^2 + \sin^2\theta
    \d\phi^2) \, .
 \ea
Now in the coordinate frame and in the gauge
$\omega^\alpha{}_\beta=0$ we calculate the nonmetricity,
$Q_{\alpha \beta} = -\frac{1}{2} \d g_{\alpha \beta}$,
 \ba
    Q_{tt} = FF' \d r \, , \quad  Q_{rr} = -GG' \d r \, , \quad   Q_{\theta \theta} = -r \d r \, , \nonumber \\
    Q_{\phi \phi} = -r \sin^2\theta \d r - r^2 \sin\theta \cos\theta \d \theta \, , \quad \mbox{others} = 0 \, ,
 \ea
where prime denotes derivative with respect to $r$. Then we pass
to the orthonormal frame
 \ba
    e^0 = F \d t \, , \quad  e^1 = G \d r \, , \quad   e^2 = r \d \theta \, ,
    \quad e^3 = r \sin\theta  \d \phi \, .
 \ea
That gives the chance of computing the nonmetricity in the
orthonormal frame
 \ba
  Q_{00} = \frac{F'}{FG} e^1 \, , \quad  Q_{11} = -\frac{G'}{G^2} e^1 \, ,
  \quad  Q_{22} = -\frac{1}{rG} e^1 \, , \quad  Q_{33} = -\frac{1}{rG} e^1 - \frac{\cot\theta}{r} e^2 \, ,
  \quad  \mbox{others} = 0 \, .
 \ea
When we substitute these findings to the field equation
(\ref{AlanDenk6}) in vacuum, that is $t_a=0$ and $\ell^a{}_b=0$,
we obtain a very complicated set of equations. Therefore we try
simplifying them firstly by choosing $G=1/F$. Then under the case
of $c_1 \neq 0$, $c_2 =c_5=-2c_1$, $c_3 = 0$, $c_4 = c_1$ a
dynamical solution is obtained
 \ba
    F^2 = 1- \frac{M}{r} + \frac{\Lambda}{6 c_1}r^2
    \label{eq:sphsym1}
 \ea
where $M$ is an integration constant. For $c_1=-1/2$ this is
exactly the Schwarzschild-de Sitter solution of GR. This result is
also a crosscheck of GR-values of $c_i$s (\ref{GRcs}). If we set
$c_1=c_2=c_4=c_5=0$, but $c_3 \neq 0$ we find another solution
 \ba
    F=a+br\, , \quad G=1
 \ea
where $a$ and $b$ are constants. These solutions are not
conformally invariant. One class of conformally invariant solution
is obtained as
 \ba
    F = \exp(-m/r) \, , \quad G= kr
 \ea
with $\Lambda=0$ and the constraints $c_1 \neq 0$, $c_2=c_3=-c_1$,
$c_4=c_5=0$, where $m$ and $k$ are arbitrary constants.

 \section{Homogeneous and isotropic cosmology}

In spite of the fact that if a model more general than the general
relativity is investigated, principally a complete model with any
kind matter such as hyperfluid, spinorial etc has to be
considered, we assume at this point that the matter lagrangian
does not depend explicitly on the connection, i.e. $\ell^a{}_b=0$.
This assumption is because we will be searching cosmological
solutions with a matter source in the form of non-spinning perfect
fluid. Otherwise, one can always drop this restriction for
spinning sources. Thus we consider the non-spinning perfect fluid
energy-momentum 3-form
 \ba
  t_a =[(\rho + p)\eta_{0a} \eta_{0b} + \eta_{ab}p]\, {}^*\!e^b  \label{eq:perfectfluid}
 \ea
where $\rho$ is matter density, $p$ is pressure. We consider the
Robertson-Walker metric in the isotropic coordinates $x^\alpha
=(t,x,y,z)$
 \ba
   g=-dt^2 + S^2(t) \frac{\d x^2 + \d y^2 + \d z^2}{(1+kr^2/4)^2}
 \ea
where $k$ is a constant, $r^2=x^2+y^2+z^2$ and $S(t)$ is the
expansion function. First, we choose such a frame and a connection
that $e^\alpha = \d x^\alpha$ and $\omega^\alpha{}_\beta =0$, then
torsion and curvature are both zero, but  $Q_{\alpha \beta} =
-\frac{1}{2} \d g_{\alpha \beta} \neq 0$,
 \ba
    Q_{xx}=Q_{yy}=Q_{zz} = -\frac{S \dot{S} \d t}{(1+kr^2/4)^2}
    + \frac{kS^2 (x\d x + y\d y +z\d z)}{2(1+kr^2/4)^3} \, , \quad \mbox{others}=0
 \ea
where dot denotes derivative with respect to $t$. Now we pass to
the orthonormal frame
 \ba
    e^0= \d t \; , \; \; e^1 = \frac{S}{1+kr^2/4} \d x \, , \quad
    e^2 = \frac{S}{1+kr^2/4} \d y \, , \quad
    e^3 = \frac{S}{1+kr^2/4} \d z
 \ea
and correspondingly calculate
 \ba
     Q_{11}=Q_{22}=Q_{33} = -\frac{\dot{S}}{S} e^0 + \frac{k}{2S} (x e^1 + y e^2 +z e^3) \, , \quad
    \mbox{others}=0 \, .
 \ea
Now we saw again that $\omega_{ab} = Q_{ab}$. We also crosschecked
GR-values of $c_i$s given by (\ref{GRcs}). Then we deal with the
case with $k=0$, because cosmic microwave background experiments
have indicated that the geometry of the Universe is spatially flat
\cite{pbernardis2000},\cite{rstompor2001}. Thus by setting $k=0$
from beginning we arrive at the equations
 \ba
     3 a \ddot{S}/S + 3 b \left(\dot{S}/S\right)^2 = -(\Lambda + \rho) \label{eq:alandenk1}\\
    (b-2a) \left[2\ddot{S}/S + \left(\dot{S}/S\right)^2\right]  = -(\Lambda - p) \label{eq:alandenk2}
 \ea
where $a=2c_4 + c_5$ and $b=c_1 + c_2 + 7c_4 + 2c_5$. Here note
that because of the assumed homogeneity, $\rho$ and $p$ are
functions of $t$ alone. Here there are two equations:
(\ref{eq:alandenk1}), (\ref{eq:alandenk2})  but three unknowns:
$S,p,\rho$. So by postulating an equation of state $p=w\rho+B$ we
close the system where $w$ and $B$ are constants. This equation of
state is a special case of the modified Chaplygin gas defined as
$p=w\rho +B / \rho^\alpha$ where $B<0$ and $0 \leq \alpha \leq 1$.
As a result we obtain a solution
 \ba
    S=\left[ K_2 e^{-nt} \left(e^{nt} - K_1 \right)^2
    \right]^{m/n}
 \ea
where $K_1$ and $K_2$ are integration constants, and
 \ba
   m= \sqrt{\frac{-B + (1+w)\Lambda}{3(2a-b -w(a+b))}} \, , \quad
   n=\frac{6m(2a-b -w(a+b))}{4a-2b - 3aw} \, . \label{cossol}
 \ea
Here we notice that since $a=0$, but $ b\neq 0$ for GR-values of
$c_i$s, although our solution is non-Einsteinian it behaves
qualitatively like Einsteinian one.

 \section{pp-Wave Solution}

We start with the metric
 \ba
    g=-dt^2 + p^2 dx^2 + q^2 dy^2 + dz^2
 \ea
where $p=p(u)$ and $q=q(u)$ are functions of the null coordinate
$u=z-t$. This metric represents the plane fronted the
gravitational waves propagating in the $z$-direction with the
speed of light. In the gauge $e^\alpha = \d x^\alpha$ and
$\omega^\alpha{}_\beta =0$ we calculate the nonmetricity,
$Q_{\alpha \beta}=-\frac{1}{2}\d g_{\alpha \beta}$
 \ba
     Q_{xx} = -p \d p \, , \quad  Q_{yy} = -q \d q \, , \quad   \mbox{others} =0\, .
 \ea
Now we pass to orthonormal frame
 \ba
    e^0 = \d t \, , \quad e^1 = p \d x \, , \quad e^2 = \d y \, , \quad e^3 = \d z \, , \nonumber \\
    Q_{11} = - \frac{p'}{p}(e^3-e^0) \, , \quad  Q_{22} = - \frac{q'}{q}(e^3-e^0) \, , \quad   \mbox{others} =0\, ,
 \ea
where prime denotes derivative with respect to $u$. When we insert
these components of nonmetricity to the Eqn(\ref{AlanDenk6}) in
the vacuum we obtain $\Lambda=0$. Then the set of equations turns
out to be
 \ba
    c_5pq (pq'' + q p'') + 2(c_1+c_2+c_4) [(pq')^2 + (qp')^2] +
    2(2c_4 +c_5) pq p'q' =0
 \ea
Here we checked that GR-equivalent values of $c_i$s gives the
Einstein equation, namely $p''/p + q''/q=0$. There are various
solutions to this equation in the literature, for example
$p=\cos(a_1u +a_2)$ and $q=\cosh(a_1u+a_3)$ where $a_i$s are
arbitrary constants. On the other hand if $c_1+c_2=0$, then
$pq=constant$ is a non-Einsteinian solution. Consequently we see
that STPG predicts gravitational waves as well.

 \section{A Dirac field coupling to STPG}

Ordinary matter is made up by fermionic particles. Therefore, we
investigate a spin-$1/2$ particle in STPG.
We, however, emphasise that since we are in a symmetric teleparallel geometry, not in the RC,
the common knowledge of spinor being the source of torsion is not valid here.
Instead, all kinds of matter source nonmetricity and also the gravitation is asribed to the nonmetricity of the spacetime.

We are using the formalism of Clifford algebra $\mathcal{C}\ell_{1,3}$-valued
exterior forms. The $\mathcal{C}\ell_{1,3}$ algebra is generated
by the relation among the orthonormal basis
$\{\gamma_0,\gamma_1,\gamma_2,\gamma_3\}$
 \begin{equation}
    \gamma_{(a} \gamma_{b)} = \eta_{ab} \, .
 \end{equation}
One particular representation of the $\gamma^a$'s is given by the
following Dirac matrices
 \begin{eqnarray}
   \gamma_0 = i\left(\begin{array}{cc}
                -I & 0 \\
                0 & I
              \end{array}\right) \, , \;
   \gamma_1 = i\left(\begin{array}{cc}
                0 & \sigma^1 \\
                -\sigma^1 & 0
              \end{array}\right) \, , \;
   \gamma_2 = i\left(\begin{array}{cc}
                0 & \sigma^2 \\
                -\sigma^2 & 0
              \end{array}\right) \, , \;
  \gamma_3 = i\left(\begin{array}{cc}
                0 & \sigma^3 \\
                -\sigma^3 & 0
              \end{array}\right)
 \end{eqnarray}
where $\sigma^1,\sigma^2,\sigma^3$ are the Pauli matrices. In this
case a Dirac spinor $\Psi$ can be represented by a 4-component
column matrix. Thus we write explicitly the covariant exterior
derivative of $\Psi$ via the Kosmann lift \cite{madak2010},\cite{madak2006},\cite{ykosmann1966}
 \ba
  \D\Psi = d\Psi + \frac{1}{4} \omega^{ab} (\eta_{ab} + 2\sigma_{ab} )\Psi \label{covderspnr}
 \ea
where $\sigma_{ab}:= \frac{1}{2}\gamma_{[a} \gamma_{b]}$ are the
generators of the Lorentz group.
Correspondingly we calculate $\D^2\Psi$
 \ba
     \D(\D{\Psi}) = \frac{1}{4} R^{ab} (\eta_{ab}+2\sigma_{ab})\Psi \, .
 \ea
Here let us note that in teleparallel geometries as the curvature
of cotangent bundle, i.e. $R^a{}_b$, is zero, the curvature of the
spinor bundle, i.e. $\frac{1}{4} R^{ab} (\eta_{ab}+2\sigma_{ab})$,
is zero as well. Some relations of the Dirac matrices are
 \ba
 \begin{array}{rcl}
   \sigma_{ab}\gamma_c & =& \gamma_c \sigma_{ab}+ \eta_{bc} \gamma_a - \eta_{ac} \gamma_b \,,\\
   \sigma_{ab} \gamma_c + \gamma_c \sigma_{ab} &=& - \epsilon_{abcd} \gamma^d \gamma_5 \,,\\
   \gamma_c \sigma_{ab} &=& \frac{1}{2} \eta_{ac} \gamma_b - \frac{1}{2} \eta_{bc} \gamma_a - \frac{1}{2}\epsilon_{abcd} \gamma^d \gamma_5 \, , \\
   {}[\sigma_{ab}, \sigma_{cd}] &=& -\eta_{ac} \sigma_{bd} - \eta_{bd} \sigma_{ac} + \eta_{ad} \sigma_{bc} + \eta_{bc} \sigma_{ad} \, ,
 \end{array} \label{dirmatrixprop}
 \ea
where $\gamma_5 := \gamma_0 \gamma_1 \gamma_2 \gamma_3$. In order
to obtain the Bjorken-Drell conventions \cite{Bjorken1964} one has
to replace $\gamma^a \rightarrow -i \gamma^a$ and $\gamma_5
\rightarrow i \gamma_5$. Now we take the minimally coupled Dirac
lagrangian given by the {\it hermitian} 4-form
 \begin{equation}
 K = \frac{i}{2}\left(\overline{\Psi}\;{}^*\!\gamma \wedge \D\Psi +
 \D\overline{\Psi}\wedge{}^*\!\gamma\,\Psi\right)
            + im \,\overline{\Psi}\Psi \, {}^*\!1 \, , \label{DirLag}
 \end{equation}
where $\gamma:=\gamma_a e^a$ is $\mathcal{C}\ell_{1,3}$-valued
1-form and $\overline{\Psi}$ is the Dirac adjoint spinor,
$\overline{\Psi}:= \Psi^\dag \gamma_0$. The coframe $e^a$
necessarily occurs in the Dirac Lagrangian, even in special
relativity. The hermiticity of the lagrangian (\ref{DirLag}) leads
to a charge current which admits the usual probabilistic
interpretation. Then for a Dirac field we calculate the
energy-momentum 3-form
 \ba
    t_a = \frac{i}{2} \left[ \overline{\Psi} \, {}^*\!(\gamma \w e_a) \w \D\Psi
     - \D \overline{\Psi} \w {}^*\!(\gamma \w e_a)\Psi \right] + i m
     \overline{\Psi} \Psi \, {}^*\!e_a \, ,
 \ea
the angular momentum 3-form
 \ba
    \ell^a{}_b = \frac{i}{4}\left( \overline{\Psi} \gamma \gamma_5 \Psi \right)
    \w e^a{}_b \, .
 \ea
The matter field equation $\delta K/\delta \overline{\Psi}=0$
yields the Dirac equation in the most general non-Riemannian
spacetime with $Q_{ab}\neq 0$, $T^a \neq 0$, $ R^a{}_b \neq 0$;
 \ba
   {}^*\gamma \w \left(\D - \frac{1}{2}T - Q\right)\Psi + \gamma_{(a}
   {}^*\!e_{b)} \w Q^{ab}\Psi + m \Psi \, {}^*\!1 =0
 \ea
where $T=\iota_aT^a$ which is zero in STPG. So, in the gauge,
$e^\alpha = \d x^\alpha$ and $\omega^\alpha{}_\beta=0$, the Dirac
equation turns out to be
 \ba
   {}^*\gamma \w \left(\d - \frac{1}{2}Q \right)\Psi + \gamma_{(a}
   {}^*\!e_{b)} \w Q^{ab}\Psi + m \Psi \, {}^*\!1 =0 \, .
 \ea

Before closing the section we want to remark on the world line of a spinning test particle.
Although the Mathisson-Papapetrou (MP) equations \cite{mathisson1937},\cite{papapetrou1951} are the first and successful attempts on this route,
it is of interest to obtain alternative formulations for various reasons, see \cite{nomura1991},\cite{leclerc2005} and references therein.
For example, as in Ref.\cite{nomura1991} the authors apply the MP method based crucially on the matter current densities
to spinning test particles moving in the RC spacetime,
in \cite{leclerc2005} the author shows that, starting with a simple lagrangian, one can derive equations of world line of spinning test particles
without explicitly referring to spin and stress energy densities in the RC geometry.
The case with nonmetricity has not been investigated in the literature, at least to our knowledge. Nevertheless since we know that point particles can be
obtained by composing spinor particles, we have to note that there may be inconsistency between definitions (\ref{partrans}) and (\ref{covderspnr}).
The first simple suggestion to remedy contradictory may be to add a linear combination of $(\iota_a Q^a{}_b)e^b \Psi$ and $Q \Psi$ to
the exterior covariant derivative of the spinor (\ref{covderspnr}).
Thus, the equations of motion for spinning particles in a non-Riemannian spacetime with nonmetricity is an original problem
which is postponed as a future research project.

 \section{Concluding Remarks}\label{sec:result}

In this paper we studied the general symmetric teleparallel
gravity model within the framework of the generic non-Riemannian
formulation. A similar but narrow-scoped work was performed in
\cite{madak2006}. Generalizing the previous work in which we had
guessed convenient coframe and full connection, we here prescribed
a gauge $\omega^\alpha{}_\beta=0$ in the coordinate frame
$e^\alpha = \d x^\alpha$ for writing down the nonmetricity
directly from a metric ansatz, then obtained new exact solutions
with the help of computer and finally considered the minimal Dirac
spinor coupling to STPG in a consistent way.

We described a peculiar rule for parallel transportation of a
tangent vector, see equation (\ref{partrans}). Thus with a special
choice of parameters in our gauge, i.e. $e^\alpha = \d x^\alpha$
and $\omega^\alpha{}_\beta=0$, the autoparallel curves of the
symmetric teleparallel geometry coincide with those of the
Riemannian geometry. We also showed that certain values of the
parameters $c_i$ correspond GR at the level of lagrangian and
solutions. It is important to know them because the GR-equivalent
STPG, for obvious reasons, is satisfactorily supported by
observations. Since the first most important motivation for a new
theory of gravity was the quantization problems of GR, and since
the conformally invariant theories predict better UV behavior
\cite{tanhayi2012}, we investigated that type of solution which is
non-Einsteinian. But whether our scale-invariant solution really
makes sense at the quantum level is an open issue. We could find
three classes of the spherically symmetric static solutions. Two
of them are non-Einsteinian. Nevertheless there may be more
solutions of this type. The second most important motivation for
looking for a new theory of gravity was the observational data on
cosmic speed-up. Correspondingly, we investigated a cosmological
solution for a non-spinning perfect fluid source with assumption
of the modified Chaplygin gas. Although it is non-Einsteinian
(since $b \neq 0$ in equation (\ref{cossol})), it qualitatively
behaves like Einsteinian solution with a cosmological constant.
Thus, our cosmological solution can predict accelerating universe
by some suitable adjustments of coupling parameters. Furthermore
since apart from Einsteinian solution we found a non-Einsteinian
pp-wave solution, we conclude that STPG predicts the existence of
gravitational waves. In this context, we want to remark that an
intriguing physical quantity that may be associated with such
waves is the helicity of the gravitational field. We postpone a
detailed analysis on this issue for a future work. Finally because
of the fact that matter is mainly made up by fermions it was
indispensable to elaborate consistently on a minimal Dirac spinor
coupling to STPG in order to make our model self-contained as much
as possible. Thus we wrote down naively the Dirac equation in the
most general non-Riemannian geometry and then reduced it to the
symmetric teleparallel geometry in our gauge.

\section*{Acknowledgment}

MA thanks T Dereli for the stimulating discussions, F W Hehl for
the fruitful correspondences and B Tekin for the valuable
conversations especially on Section \ref{sec.conform}.

\end{document}